\documentclass[11pt, a4paper,  notitlepage]{extarticle}

\usepackage[left=3.5cm, right=1.5cm, top=2.5cm, bottom=2.5cm, bindingoffset=0cm]{geometry}
\usepackage{graphicx}
\usepackage{mathtools}
\usepackage[affil-it]{authblk}
\usepackage{url}
\usepackage{amsmath}
\usepackage{amssymb}
\usepackage{multicol}
\usepackage{float}
\usepackage[section]{placeins}
\usepackage{hyperref}
\begin{document}
\title{Regression Approach for Modeling COVID-19 Spread and Its Impact On Stock Market}
\author{Bohdan M. Pavlyshenko \\  
\small{ email:b.pavlyshenko@gmail.com, LinkedIn:https://www.linkedin.com/in/bpavlyshenko/}}
\maketitle
\section{Abstract}
The paper studies different regression approaches for modeling COVID-19 spread  and its impact on the stock market. The logistic curve model was used with Bayesian regression for predictive analytics of the coronavirus spread. 
The impact of  COVID-19 was studied using regression approach and compared to other crises influence. 
In practical analytics, it is important to find the maximum of coronavirus cases per day, this point means  the estimated half time of coronavirus spread in the region under investigation. 
	The obtained results show that different crises with different reasons have different impact on the same stocks. It is important to analyze their impact separately. Bayesian inference makes it possible to analyze the uncertainty of crisis impacts.

Keywords: COVID-19, coronavirus, regression, predicitive analyticvs, stock market,  Bayesian inference

 \FloatBarrier
\section{Bayesian Model for COVID-19 Spread Prediction}
At present time, there are different methods, approaches, data sets for for modeling  COVID-19 spread  ~\cite{ref2,ref3,ref4,ref5,ref6,ref7}.
For the predictive analytics of COVID-19 spread, we used a logistic curve model. 
Such model is very popular nowadays. 
To estimate model parameters, we  used Bayesian regression \cite{kruschke2014doing,gelman2013bayesian, carpenter2017stan}.
 This approach allows us to receive a posterior distribution of model parameters using conditional likelihood and prior distribution. 
  In the Bayesian inference, we can use informative prior distributions which can be set up by an expert. So, the result can be considered as a compromise between 
historical data and expert opinion. It is important in the cases when we have a small number of historical data. 
Probabilistic approach makes it possible to receive the probability density function for the target variable. 
Logistic curve model with Bayesian regression approach can be written as follows:
 \begin{equation}
 \begin{split}
 &n \sim \mathcal{N}(\mu, \sigma) \\
 &\mu=\frac{\alpha}{1+exp(-\beta(t-t_0))}10^5 \\
 &t=t_{weeks}(Date-Date_0)
 \end{split}
\end{equation}
where $Date_0$ is a start day for observations in the historical data set, it is measured in weeks. 
The data for the analysis were taken from ~\cite{ref3}.
 $\alpha$ parameter denotes maximum cases of coronavirus, $\beta$ parameter is an empirical coefficient which denotes the rate of coronavirus spread. 

For solving Bayesian models, numerical Monte-Carlo methods are used. Gibbs and Hamiltonian sampling are the popular methods of finding posterior distributions for the parameters of probabilistic mode
~\cite{kruschke2014doing,gelman2013bayesian, carpenter2017stan}.
Bayesian inference makes it possible to obtain probability density functions for model parameters and estimate the uncertainty that is important in risk assessment analytics. In Bayesian regression approach, we can take into account expert opinions via information prior distribution. For Bayesian inference calculations, we used pystan package for Stan platform for statistical modeling \cite{carpenter2017stan}.
Figure ~\ref{fig1} shows the box plots for $\beta$ parameters of coronavirus spread model for different countries.
\begin{figure}[htb]
\center
\includegraphics[width=0.7\linewidth]{imgs/beta_boxplots.png}
\caption{Box plots for beta coefficients of coronavirus spread model for different countries}
\label{fig1}
\end{figure}

\begin{figure}[htb]
\center
\includegraphics[width=0.85\linewidth]{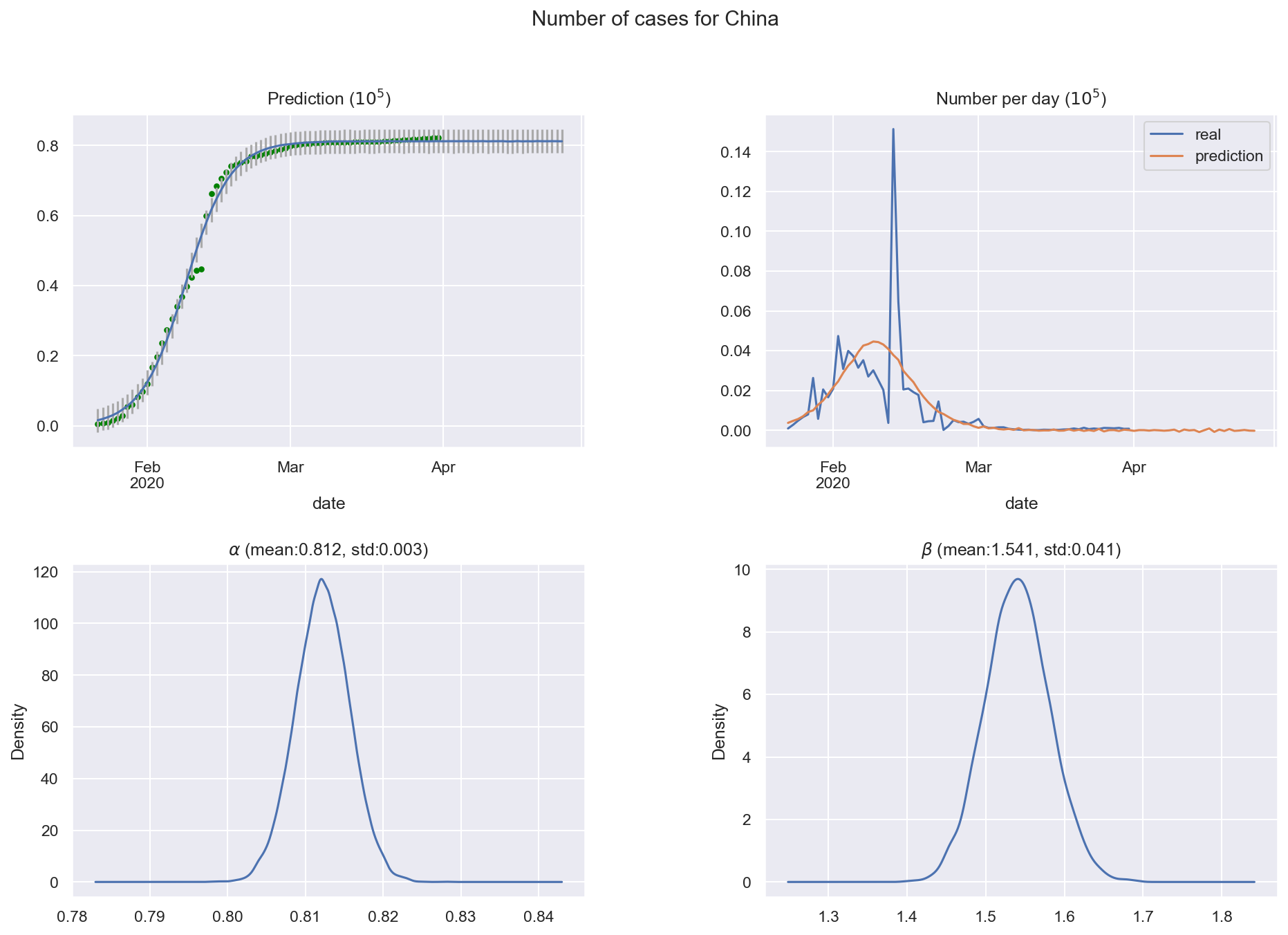}
\caption{Modeling of COVID-19 spread  for China}
\label{fig2}
\end{figure}

\begin{figure}[htb]
\center
\includegraphics[width=0.85\linewidth]{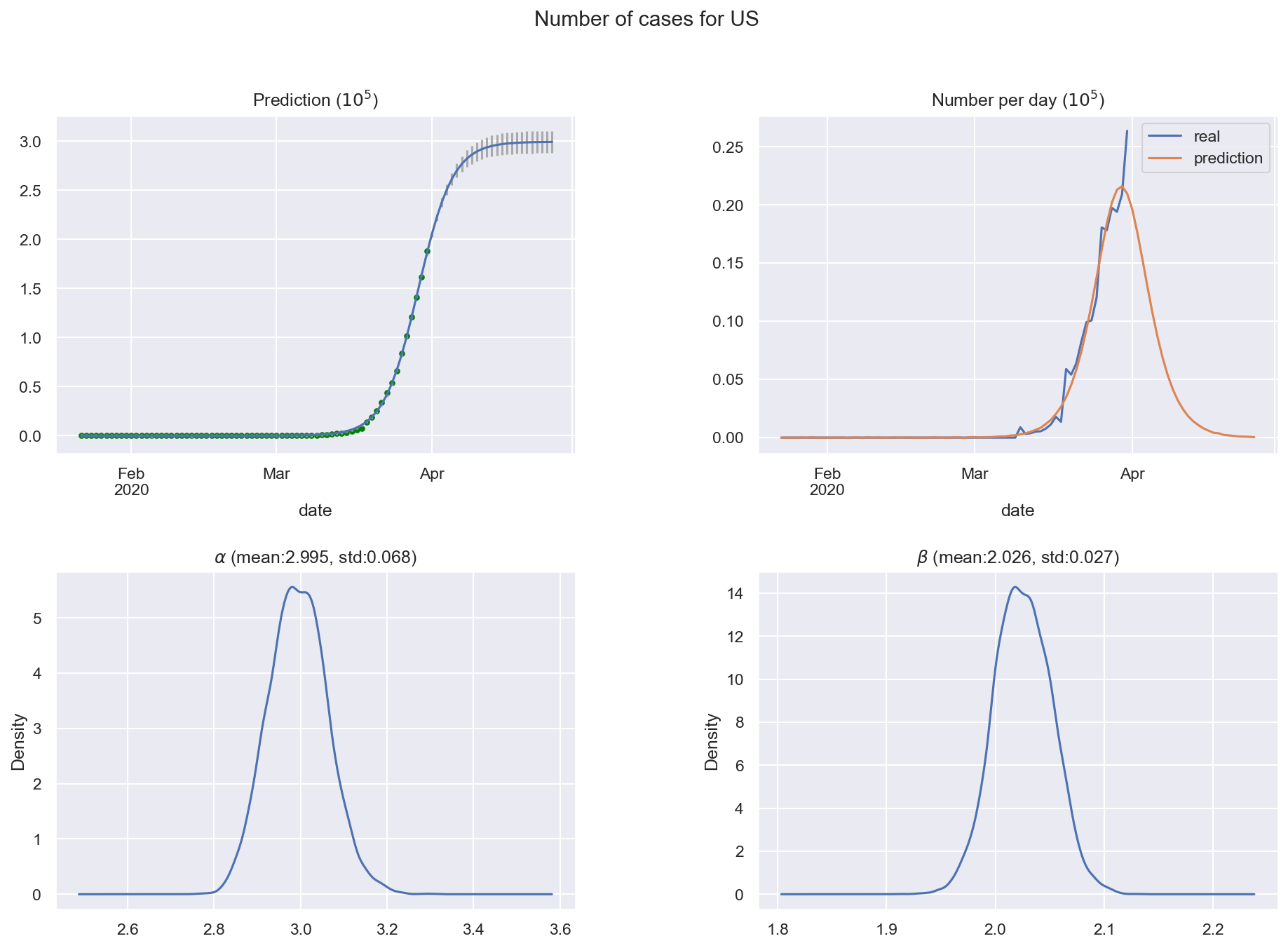}
\caption{Modeling of COVID-19 spread  for USA}
\label{fig3}
\end{figure}

\begin{figure}[htb]
\center
\includegraphics[width=0.85\linewidth]{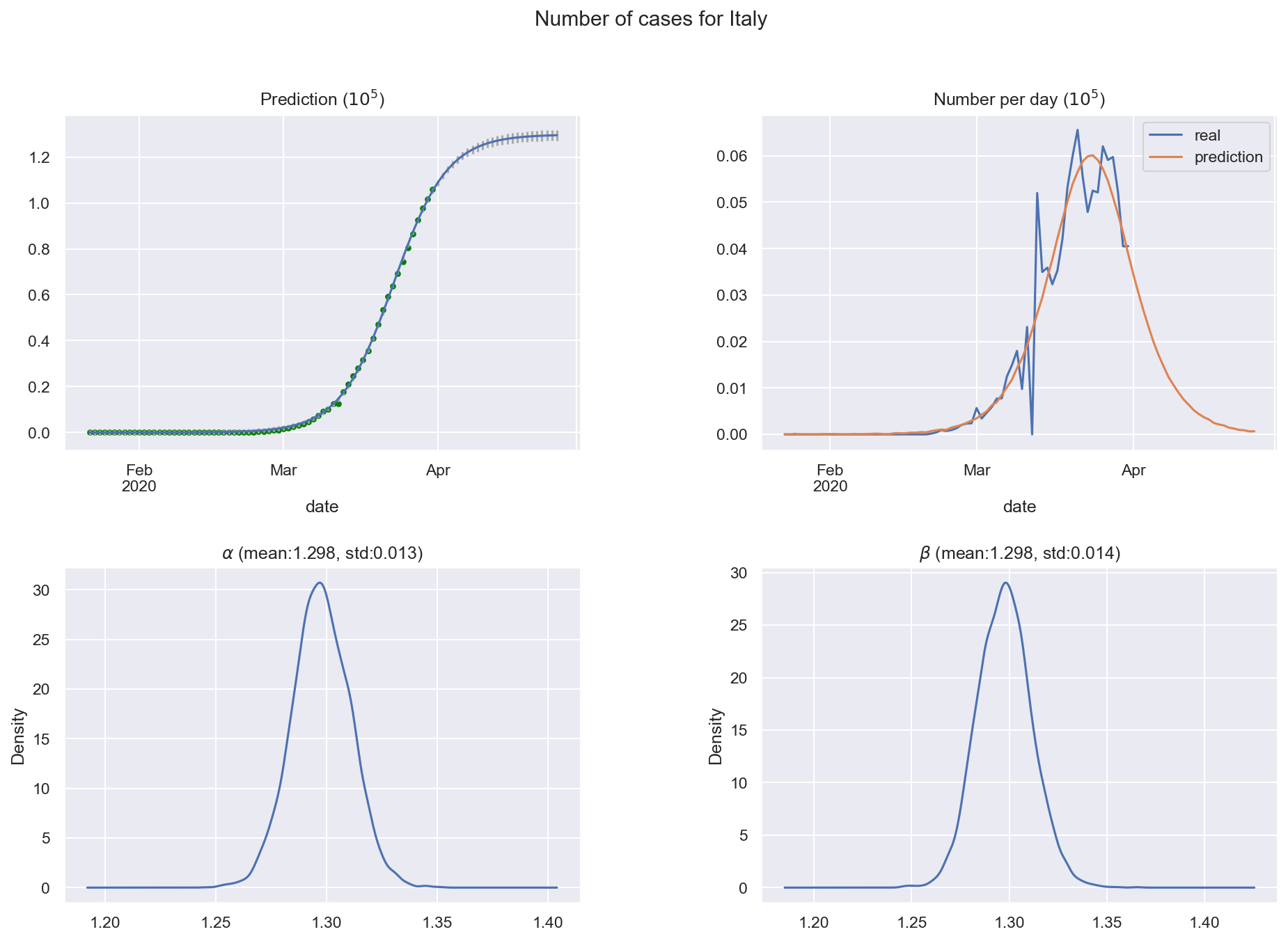}
\caption{Modeling of COVID-19 spread  for Italy}
\label{fig4}
\end{figure}

\begin{figure}[htb]
\center
\includegraphics[width=0.85\linewidth]{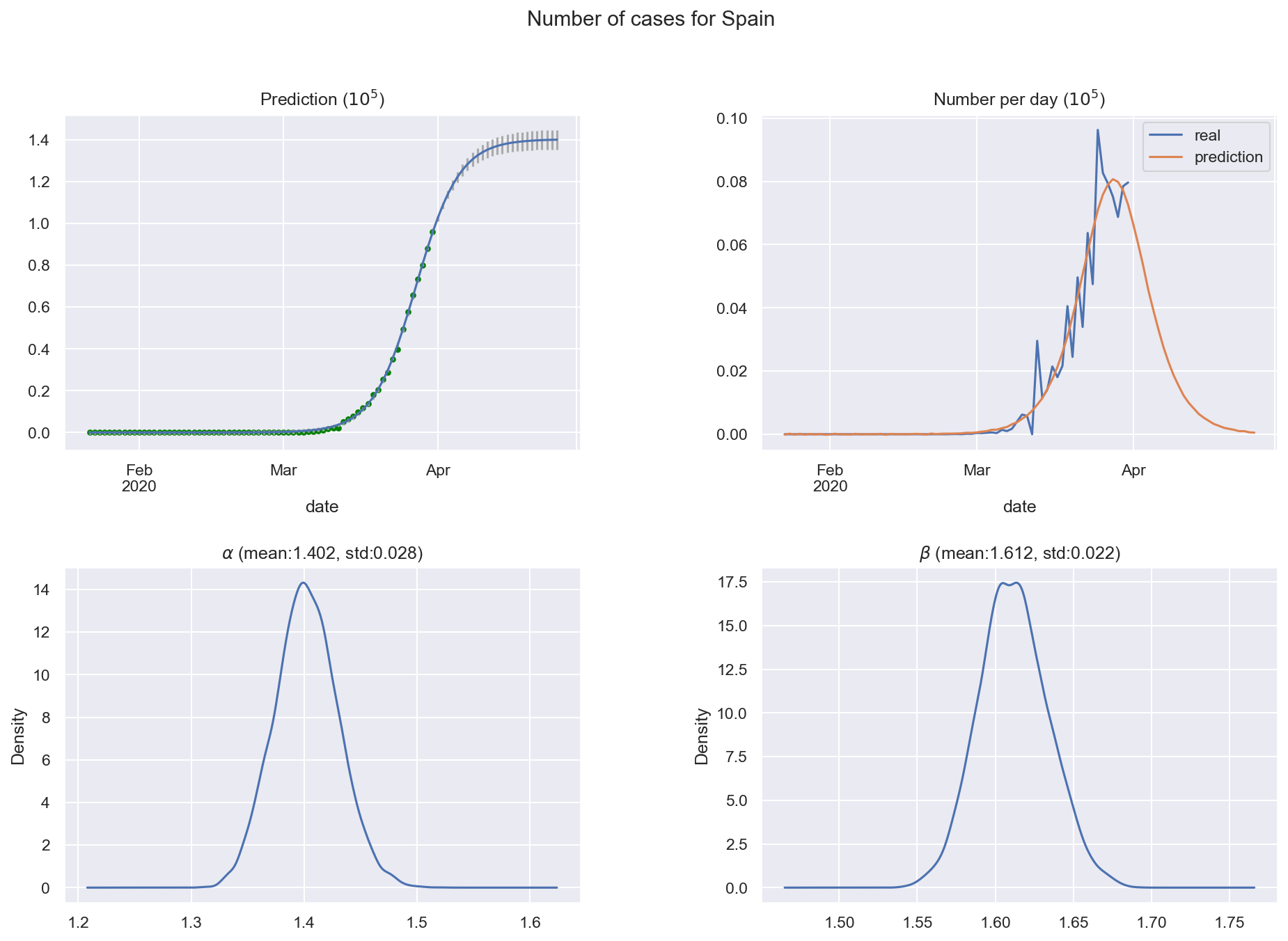}
\caption{Modeling of COVID-19 spread  for Spain}
\label{fig5}
\end{figure}

\begin{figure}[htb]
\center
\includegraphics[width=0.85\linewidth]{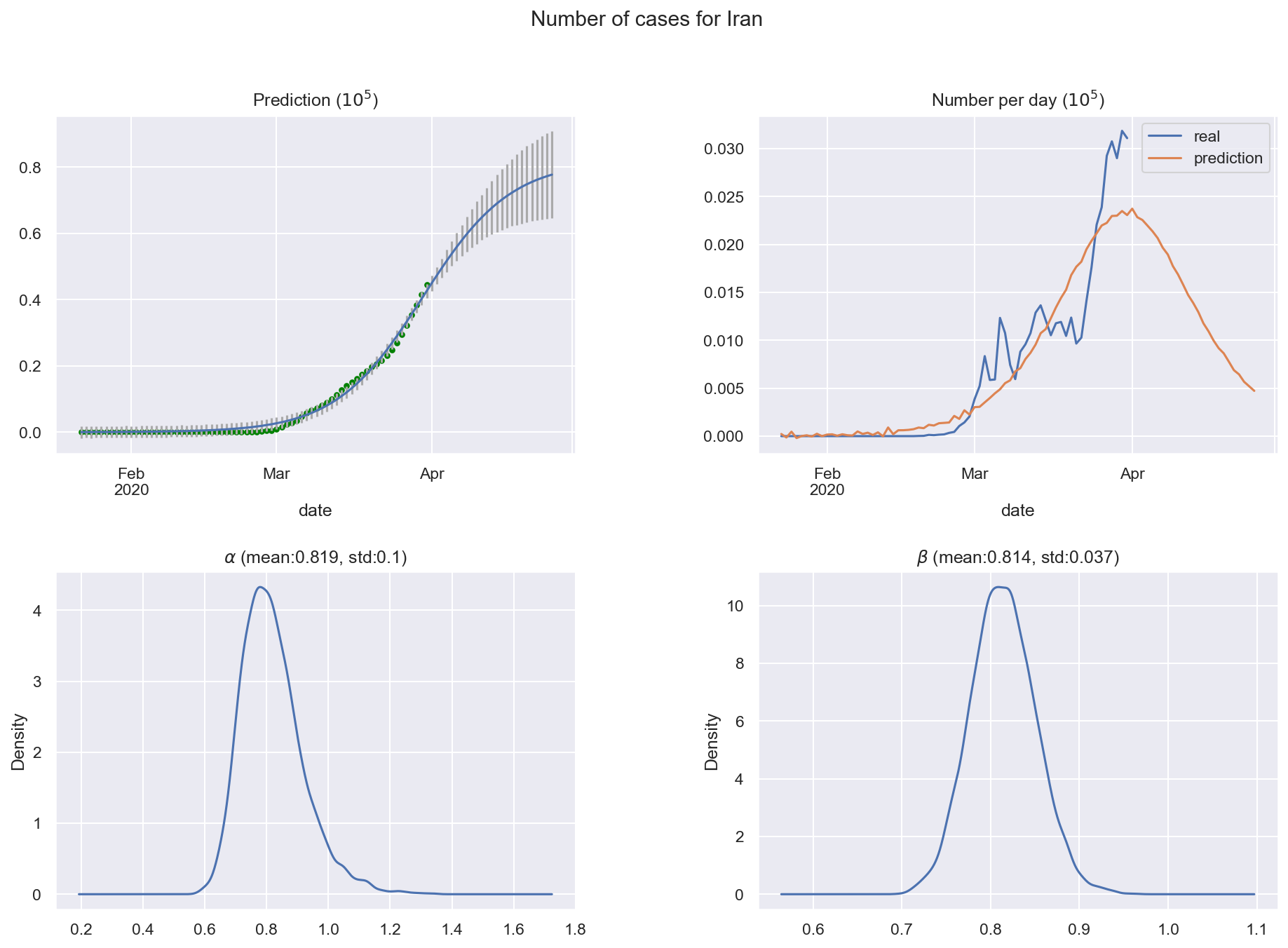}
\caption{Modeling of COVID-19 spread  for Iran}
\label{fig6}
\end{figure}

\begin{figure}[htb]
\center
\includegraphics[width=0.85\linewidth]{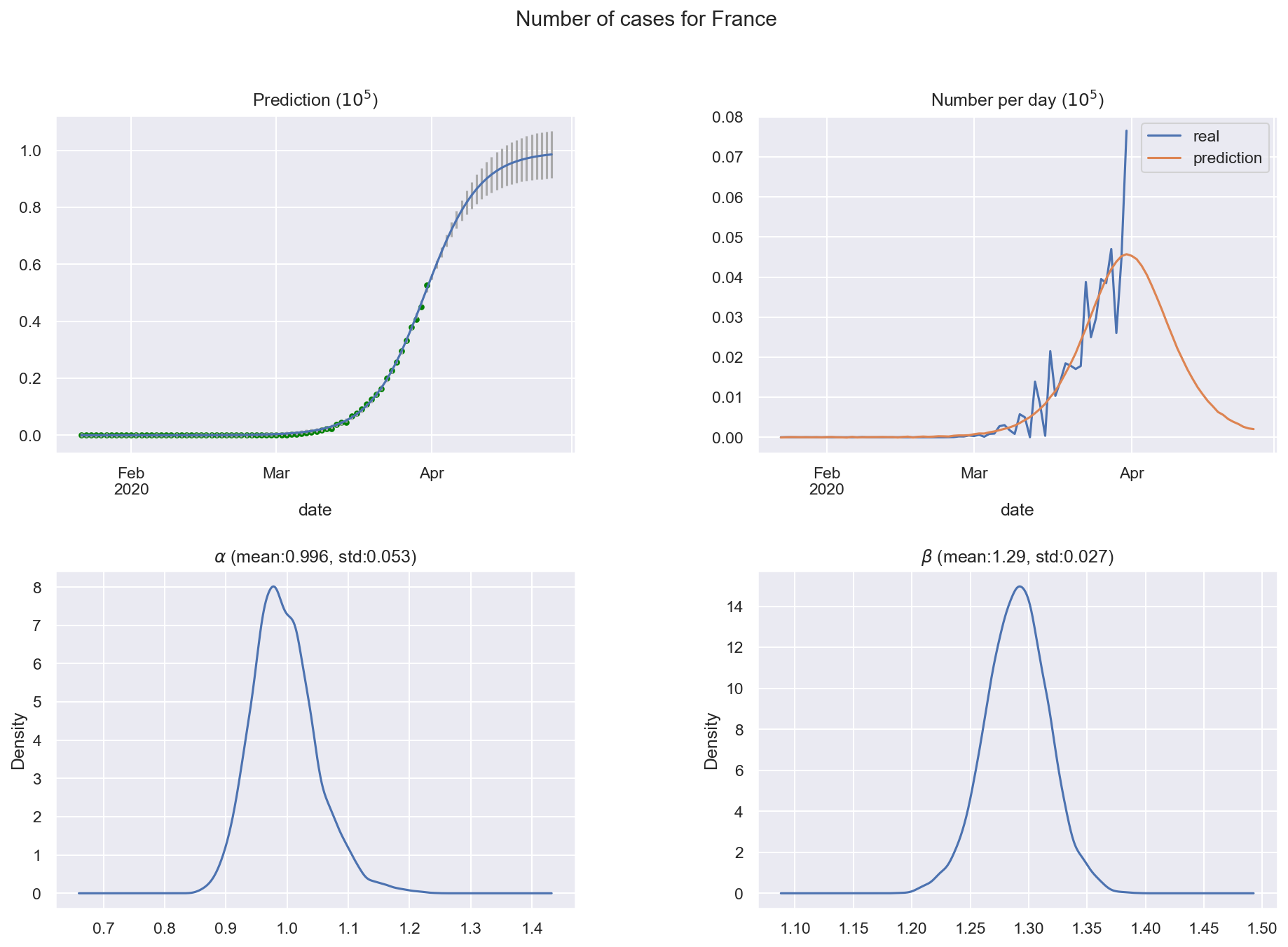}
\caption{Modeling of COVID-19 spread  for France}
\label{fig7}
\end{figure}

\begin{figure}[htb]
\center
\includegraphics[width=0.75\linewidth]{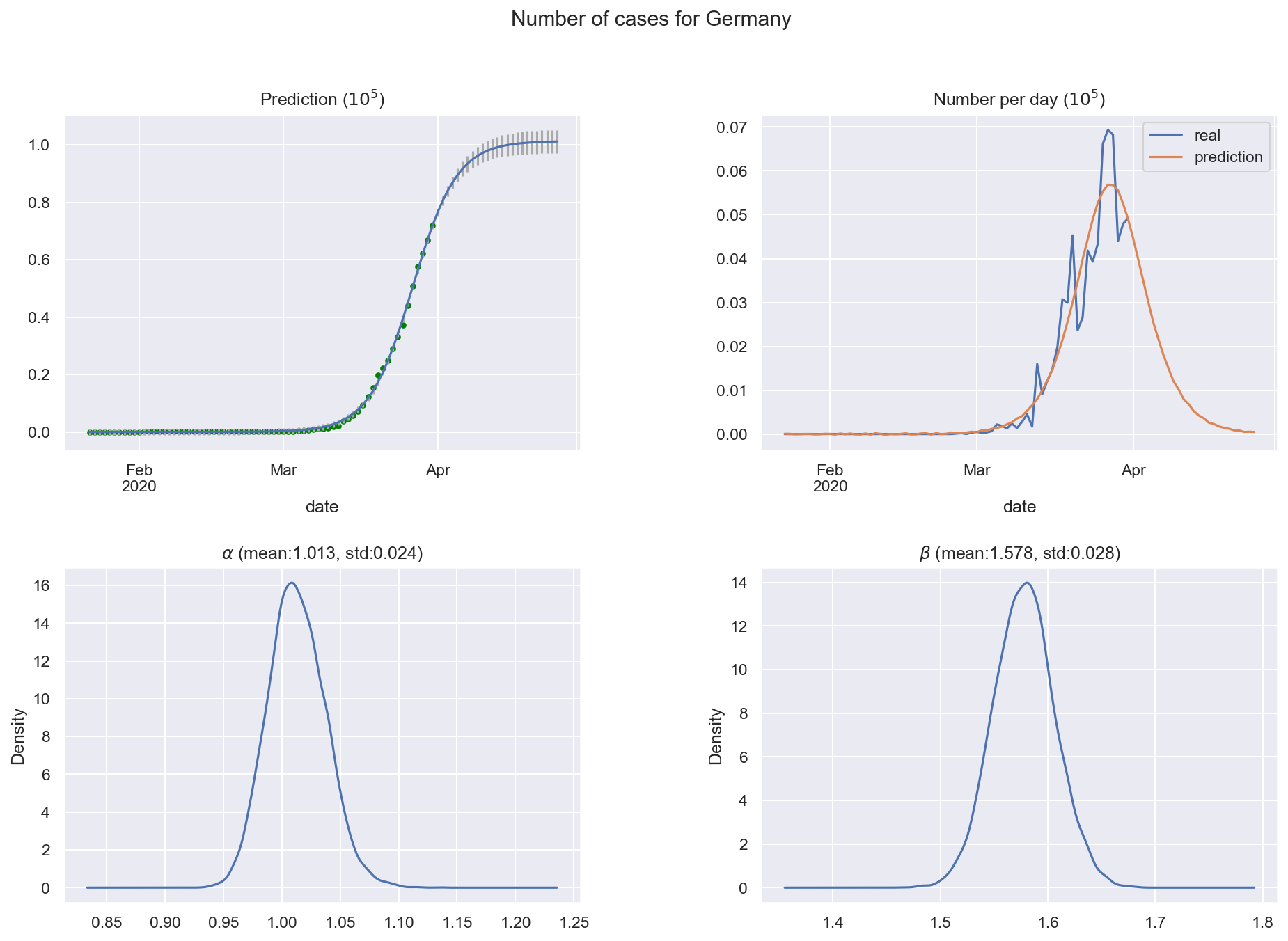}
\caption{Modeling of COVID-19 spread  for Germany}
\label{fig8}
\end{figure}

Figures~\ref{fig2},~\ref{fig3},~\ref{fig4},~\ref{fig5},~\ref{fig6},~\ref{fig7},~\ref{fig8}  show the predictions for  coronavirus spread cases  using current historical data. In  practical analytics, it is important to find the maximum of coronavirus cases per day, this point means the estimated half time of coronavirus spread in the region under investigation. New historical data will correct the distributions for model parameters and forecasting results. 
The  'Bayesian Model for COVID-19 spread Prediction' package can be loaded at ~\cite{ref8} for free use. 
 \FloatBarrier

\FloatBarrier
\section{COVID-19 Impact on the Stock Market}
	Coronavirus outbreak has a huge impact on the stock market. It is very important , e.g. for forming stable portfolios, to understand how different crises impact stock prices. 
	We are going to consider the impact of coronavirus crisis on stocks and compare it to the crisis of 2008 and market downturn of 2018. For this, we can use the regression approach using OLS regression and Bayesian regression. Bayesian inference makes it possible to obtain probability density functions for coefficients of the factors under investigation and estimate the uncertainty that is important in the risk assessment analytics. In Bayesian regression approach,  we can analyze extreme target variable values using non gaussian distributions with fat tails.
	We took the following time periods for each of crises -  crisis\_2008: [2008-01-01,2009-01-31], down\_turn\_2018: [2018-10-01,2019-01-03], coronavirus: [2020-02-18,2020-03-25].  For each of above mentioned crises, we created a regression variable which is equal to 1 in the crisis time period and 0 in other cases.
	Figure~\ref{fig10} shows the time series for S\&P500 composite index:

\begin{figure}[htb]
\center
\includegraphics[width=0.75\linewidth]{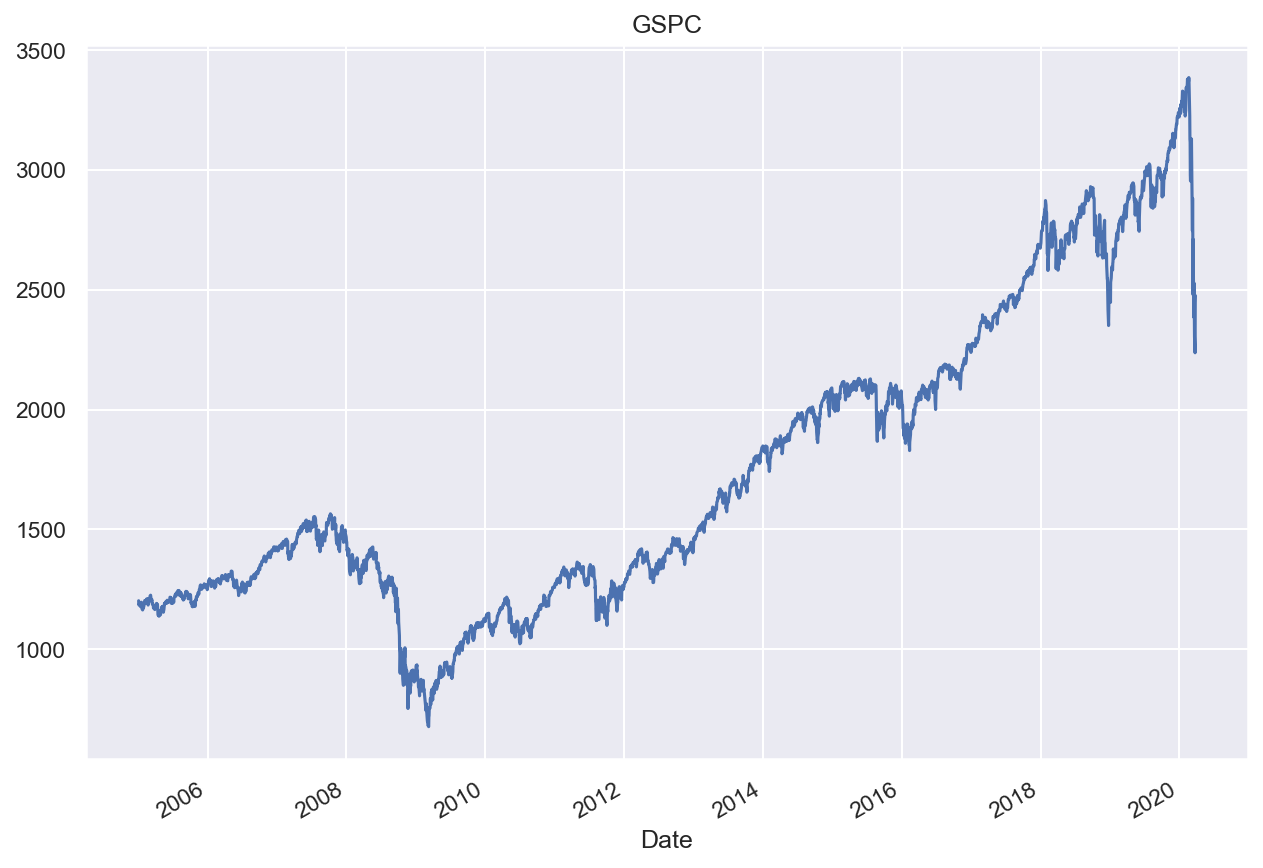}
\caption{Time series for S\&P500 composite index}
\label{fig10}
\end{figure}

\begin{figure}[htb]
\center
\includegraphics[width=0.5\linewidth]{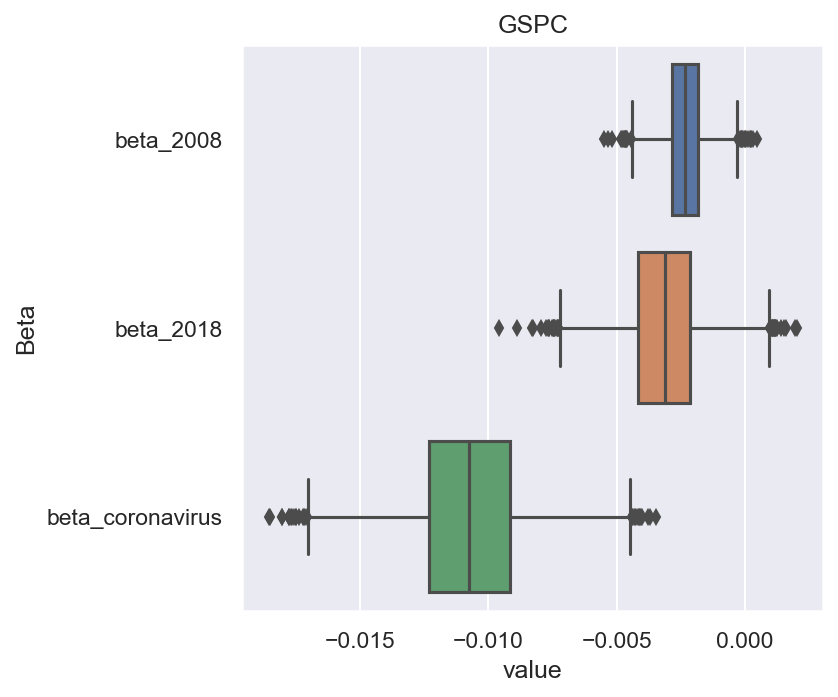}
\caption{Box plots of impact weights of each crisis on  S\&P composite index}
\label{fig11}
\end{figure}

\begin{figure}[htb]
\center
\includegraphics[width=0.5\linewidth]{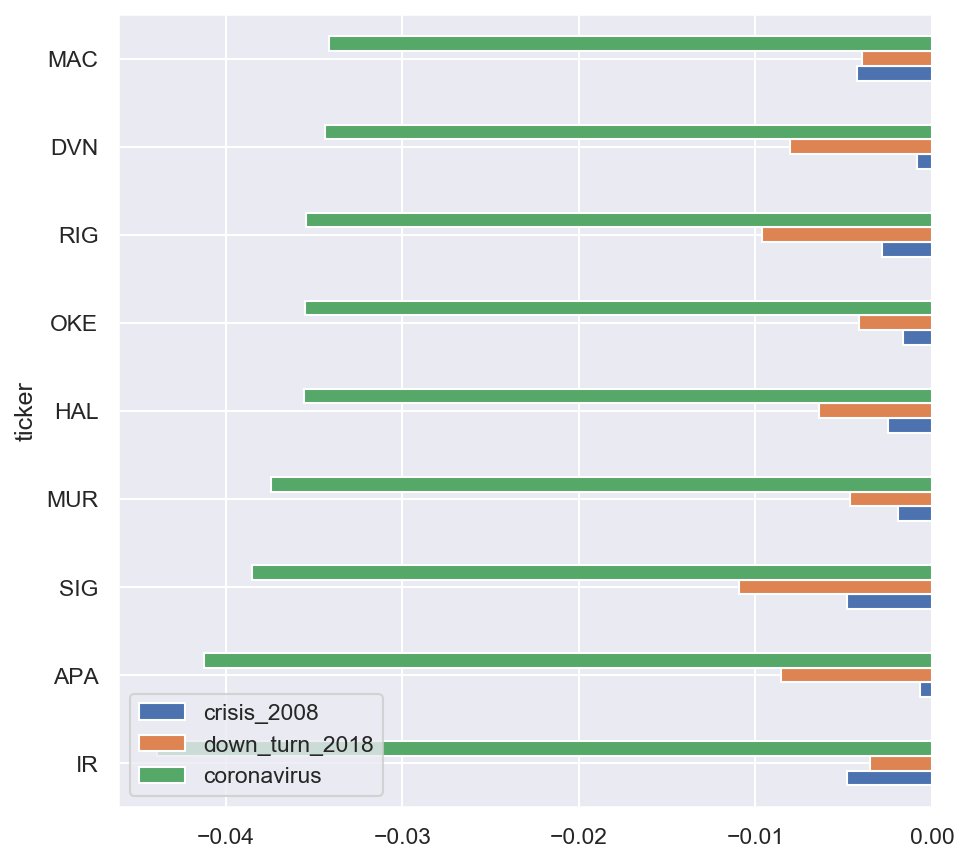}
\caption{Top negative price returns in COVID-19 crises}
\label{fig12}
\end{figure}

\begin{figure}[htb]
\center
\includegraphics[width=0.5\linewidth]{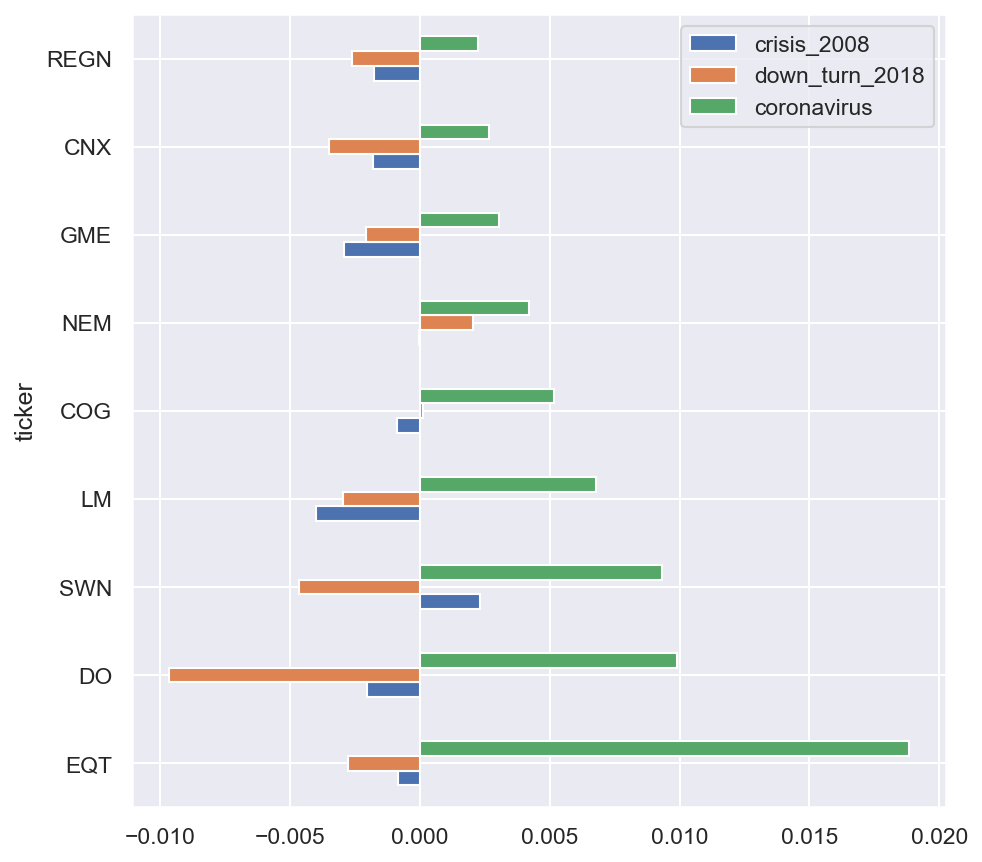}
\caption{Tickers with positive price returns in COVID-19 crises}
\label{fig13}
\end{figure}

\begin{figure}[htb]
\center
\includegraphics[width=0.5\linewidth]{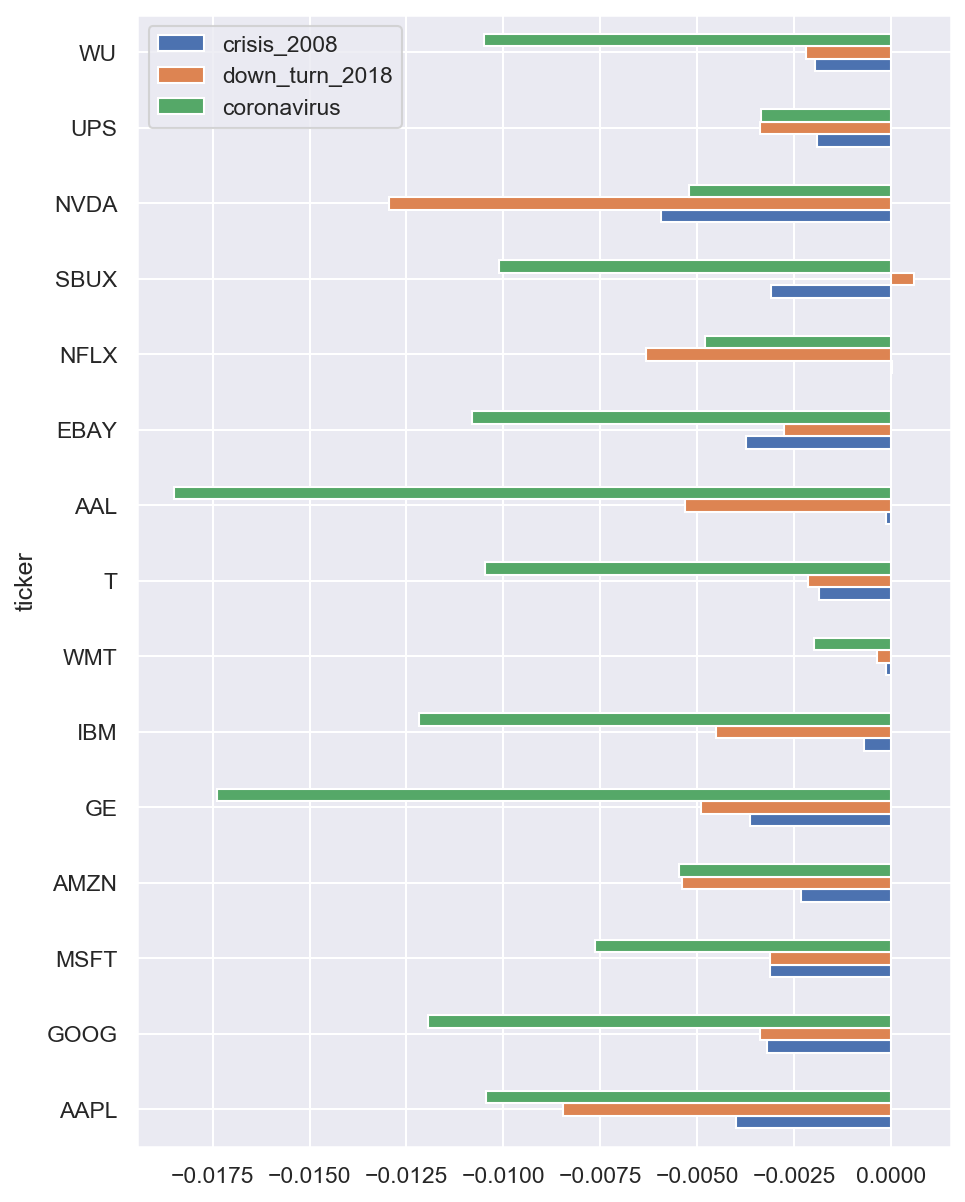}
\caption{The weights for different crises for arbitrarily chosen stocks}
\label{fig14}
\end{figure}

\begin{figure}[htb]
\center
\includegraphics[width=0.85\linewidth]{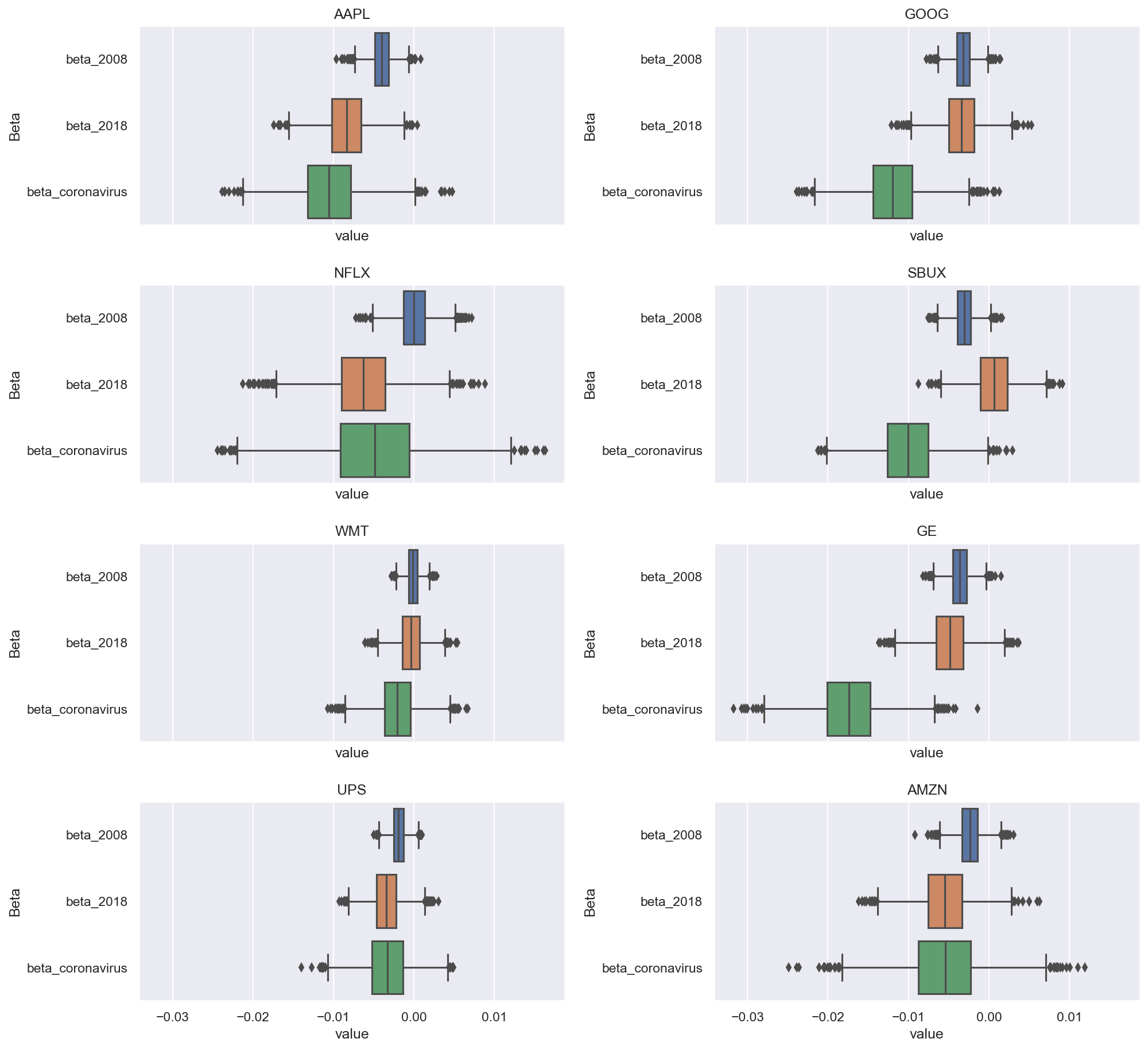}
\caption{Box plots for crisis weights for different stocks.}
\label{fig15}
\end{figure}

	As a target variable, we consider the daily price return. Knowing the daily price return, changes in crises periods, one can estimate the ability of investors to understand trends and recalculate portfolios.
	These results were received using Bayesian inference. For Bayesian inference calculations, we used Python pystan package for Stan platform for statistical modeling \cite{carpenter2017stan}.
	Figure~\ref{fig11} shows the box plots of impact weights of each crisis on  S\&P500 composite index. 
	The wider box for coronavirus weight can be caused by shorter time period of investigation comparing with other crises and consequently larger uncertainty.
	For our investigations, we took a random set of tickers from S\&P list.
	Figure~\ref{fig12} shows top negative price returns in coronavirus crises.
	Figure~\ref{fig13} shows the tickers with positive price return in coronavirus crisis. 
	Figure~\ref{fig14} shows the weights for different crises for arbitrarily chosen stocks. 
	We calculated the distributions for crises weights using Bayesian inference.
	Figure~\ref{fig15} shows the box plots for crisis weights for different stocks.

 \FloatBarrier
\section{Conclusion}
The logistic curve model can be used with Bayesian regression for predictive analytics of coronavirus spread. 
Such a model can be effective when  the exponential growth of number of coronavirus confirmed cases 
takes place.
In practical analytics, it is important to find the maximum of coronavirus cases per day, this point means  the estimated half time of coronavirus spread in the region under investigation. New historical data will correct the distributions for model parameters and forecasting results. 
'Bayesian Model for COVID-19 spread Prediction' package can be loaded at \cite{ref8} for free use. 	

	The obtained results show that different crises with different reasons have different impact on the same stocks. It is important to analyze their impact separately. Bayesian inference makes it possible to analyze the uncertainty of crisis impacts.
	The uncertainty of crisis impact weights can be measured as a standard deviation for weight probability density functions.
	The uncertainty of coronavirus crisis is larger comparing to other crises that can be caused by shorter analysis time. Knowing the uncertainty allows making risk assessment for portfolios and other financial and business processes.
   In Bayesian regression,  we can receive a quantitative measure for the uncertainty that can be a very  useful information for experts in model selection and stacking. 
  An  expert can also set up informative prior distributions for  stacking regression  coefficients of models,
   taking into account the domain knowledge information. 

\bibliographystyle{unsrt}
\bibliography{article.bib}
\end{document}